\documentclass[12pt,epsf]{article}
\usepackage{amssymb,amsmath}
\usepackage{graphicx}

\newcommand{\be}{\begin{equation}}
\newcommand{\ee}{\end{equation}}
\newcommand{\bea}{\begin{eqnarray}}
\newcommand{\eea}{\end{eqnarray}}
\newcommand{\beas}{\begin{eqnarray*}}
\newcommand{\eeas}{\end{eqnarray*}}
\newcommand{\ba}{\begin{array}}
\newcommand{\ea}{\end{array}}

\newcommand{\nbox}{{\,\lower0.9pt\vbox{\hrule \hbox{\vrule height 0.2 cm \hskip 0.19 cm \vrule height 0.2 cm}\hrule}\,}}

\def\href#1#2{#2}
\input{tcilatex}

\begin{document}

\begin{titlepage}
\hfill
\vbox{
    \halign{#\hfil         \cr
           } 
      }  

\hbox to \hsize{{}\hss \vtop{
\hbox{MCTP-08-50}

}}

\vspace*{20mm}
\begin{center}
{\Large \bf Kac-Moody Extensions of 3-Algebras \\ and M2-branes}

\vspace*{15mm}
\vspace*{1mm}
Hai Lin{\footnote {e-mail: hailin@umich.edu}}

\vspace*{1cm}

{\it Department of Physics and Michigan Center for Theoretical Physics \\
University of Michigan, Ann Arbor, MI 48109, USA \\}

\vspace*{1cm}
\end{center}

\begin{abstract}
We study the 3-algebraic structure involved in the recently shown
M2-branes worldvolume gauge theories. We first extend an arbitrary
finite dimensional 3-algebra into an infinite dimensional
3-algebra by adding a mode number to each generator. A unique
central charge in the algebra of gauge transformations appears
naturally in this extension. We present an infinite dimensional
extended 3-algebra with a general metric and also a different
extension with a Lorentzian metric. We then study ordinary finite
dimensional 3-algebras with different signatures of the metric,
focusing on the cases with a negative eigenvalue and the cases
with a zero eigenvalue. In the latter cases we present a new
algebra, whose corresponding theory is a decoupled abelian gauge
theory together with a free theory with global gauge symmetry, and
there is no negative kinetic term from this algebra.

\end{abstract}

\end{titlepage}

\vskip 1cm

\section{Introduction}

Recently it has been shown that a 3-algebraic structure is relevant for the
supersymmetry and gauge symmetry transformations \cite{Bagger:2007jr},\cite%
{Bagger:2007vi},\cite{Gustavsson:2007vu} of the worldvolume theory of
multiple coincident M2-branes. A candidate Lagrangian description of this
theory has been found after obtaining on-shell equations of motion, arising
from demanding the on-shell closure of the supersymmetry algebra. Earlier
work on the non-propagating nature of the gauge fields in this theory
includes the conjecture in \cite{Schwarz:2004yj} in which a Chern-Simons
type self-coupling of the gauge fields was proposed to be part of the
dynamics in the multiple M2-brane theory. This type of couplings not only
does not introduce new independent degrees of freedom, but also has the
right conformal dimension in three dimensions. The 3-algebraic structure of
this theory has also been hinted at by the early study of a system of
M2-branes ending on M5 \cite{Basu:2004ed}, in which a Nambu-Poisson type
3-bracket \cite{Nambu:1973qe} played an important role and had the
ingredient for making all the transverse scalars on equal footing. The
complete Lagrangian with all the requisite symmetries and a 3-algebraic
gauge symmetry including a particular $so(4)~$example were found in the
illuminating work of \cite{Bagger:2007jr},\cite{Bagger:2007vi},\cite%
{Gustavsson:2007vu}, and the Lagrangian theory has recently been studied
from various perspectives \cite{Gustavsson:2008dy}-\cite{Hosomichi:2008jd}.

In order to better understand the nature of the algebraic structure of the
fields on the worldvolume of multiple M2-branes, in this paper, we study the
3-algebraic structure itself. We extend the 3-algebraic structure considered
in \cite{Bagger:2007jr},\cite{Bagger:2007vi},\cite{Gustavsson:2007vu},\cite%
{Gustavsson:2008dy} in two directions. In the first direction, we make
extensions of the finite dimensional 3-algebras into infinite dimensional
3-algebras by adding a mode number to each generator. This extension is
similar in spirit to the Kac-Moody extension of Lie algebras. A central
charge naturally appears on the right hand side of the algebraic relations.
We also present a different extension for the Lorentzian 3-algebras \cite%
{Gomis:2008uv},\cite{Benvenuti:2008bt},\cite{Ho:2008ei}. This extension may
be relevant to the fields on M2-branes if they are also valued in an
internal circle, which may be viewed as the boundary circle of open
membranes stretching between M2-branes. In the second direction, we explore
the finite dimensional 3-algebra, but with different signatures of the
metric for the generators, with the motivation of embedding general Lie
algebras, and present a very simple 3-algebra with a zero eigenvalue in the
metric. However, the gauge theory for this simple algebra is not very
appealing and is a decoupled abelian gauge theory together with a theory
with global symmetries of an arbitrary Lie algebra, from the point of view
of the Lagrangian.

The organization of this paper is as follows. In section \ref{infinite
dimensional algebra}, we focus on the derivation of the extensions of the
3-algebras into infinite dimensional ones with mode numbers. In section \ref%
{algebra-metric-lagrangian}, we explore 3-algebras with different metric
signatures, and related Lagrangian theories. In subsection \ref%
{algebra_metric}, after revisiting the derivation of the algebras with a
negative eigenvalue in the metric, independently obtained by \cite%
{Gomis:2008uv},\cite{Benvenuti:2008bt},\cite{Ho:2008ei},\cite{Lin:200804},
we discuss 3-algebras with a zero eigenvalue in the metric. In subsection %
\ref{lagrangian}, we study Lagrangians before contracting with the metric
and study the theory corresponding to the algebra with a zero eigenvalue in
the metric. We also emphasize the study of a Lagrangian 2-tensor which falls
in the algebra of gauge transformations. In section \ref{discussion}, we
make brief conclusions and discuss mass deformed theories and related work.

\section{Extensions and infinite dimensional algebras}

\label{infinite dimensional algebra}

\subsection{Infinite dimensional extensions with general metrics}

\label{infinite dimensional algebra_general metric}

The algebras in \cite{Bagger:2007jr},\cite{Bagger:2007vi} and \cite%
{Gustavsson:2007vu} are intimately connected with each other. As was
speculated in \cite{Gustavsson:2007vu}, the scalars and spinors on the
M2-branes may live in an algebra $\mathcal{A,}$ and the gauge fields may
live in a possibly different algebra $\mathcal{B}$. 
In the case of a conventional Yang-Mills theory, these two algebras are the
same. A natural generalized possibility for M2-branes is that these two may
not be the same. Thereby there are 3 types of 2-brackets, as formulated in
\cite{Gustavsson:2007vu}:
\begin{eqnarray}
\langle \cdot ~,~\cdot \rangle &:&~\mathcal{A\otimes A}\rightarrow \mathcal{B%
}  \label{brackets_1} \\
(\cdot ~,~\cdot ) &:&~\mathcal{B~\mathcal{\otimes ~A}}\rightarrow \mathcal{A}
\label{brackets_2} \\
\lbrack \cdot ~,~\cdot ] &:&~\mathcal{B~\mathcal{\otimes ~B}}\rightarrow
\mathcal{B}  \label{brackets_3}
\end{eqnarray}

The first bracket (\ref{brackets_1}) means that we can form an antisymmetric
product of two elements in $\mathcal{A}$ to obtain a gauge transformation.
The second bracket (\ref{brackets_2}) is a gauge transformation of the
element in $\mathcal{A}$ by the action of an element in $\mathcal{B.~}$The
third bracket (\ref{brackets_3}) means that applying two gauge
transformations is again a gauge transformation. It may be worth mentioning
that if the first product (\ref{brackets_1}) is a symmetric product instead
of an antisymmetric product, then the algebra may be viewed as a super Lie
algebra. However it is not, since the first bracket is an antisymmetric
product, and is not a symmetric product.

If we make the combined operation of the first bracket (\ref{brackets_1})
and the second bracket (\ref{brackets_2}), we obtain the 3-bracket
formulated in \cite{Bagger:2007jr},\cite{Bagger:2007vi} as well as \cite%
{Gustavsson:2007vu}:\qquad
\begin{equation}
\lbrack \cdot ~,~\cdot ~,~\cdot ]=(\langle \cdot ~,~\cdot \rangle ,~\cdot ):~%
\mathcal{A\otimes A\otimes A}\rightarrow \mathcal{A}  \label{3-bracket}
\end{equation}

In principle, we may also consider the case that some elements in $\mathcal{B%
}$ are not reached by all possible products in (\ref{brackets_1}), and
similarly the case that some elements in $\mathcal{A}$ are not reached by
all possible products in (\ref{3-bracket}). However, we will not discuss
these aspects in this section, and they will not influence the general
discussion below.

Now we can introduce the basis of the elements in $\mathcal{A}$,$\mathcal{~B~%
}$for the extension of the 3-algebra into the one with integer mode numbers:
\begin{eqnarray}
~\mathcal{A}\mathcal{:~} &&\{T_{m}^{a}\}  \label{elements_A} \\
~~\ \ \mathcal{B}\mathcal{:\mathcal{~}} &&\{V_{m}^{ab},~C_{m,n}\}
\label{elements_B}
\end{eqnarray}%
where $a,b~$are gauge indices, $m,n~$are integers, which are mode numbers.
Fixing all the modes to 0, we get back the ordinary 3-algebra with zero-mode
generators.

We postulate a realization of the algebraic relations (\ref{brackets_1}),(%
\ref{brackets_2}),(\ref{3-bracket}) as:%
\begin{eqnarray}
\left\langle T_{m}^{a},T_{n}^{b}\right\rangle &=&V_{m+n}^{ab}+h^{ab}C_{m,n}
\label{alge01} \\
(V_{m}^{ab},T_{n}^{c})~ &=&~f_{d}^{abc}T_{m+n}^{d}  \label{alge02} \\
(C_{m,n}~,T_{l}^{c}) &=&g_{mn,lp}T_{p}^{c}  \label{alge03} \\
\lbrack T_{m}^{a},T_{n}^{b},T_{l}^{c}]
&=&~f_{d}^{abc}T_{m+n+l}^{d}+h^{ab}g_{mn,lp}T_{p}^{c}+h^{bc}g_{nl,mp}T_{p}^{a}+h^{ca}g_{lm,np}T_{p}^{b}
\label{alge04}
\end{eqnarray}%
$h^{ab}$ and $f_{d}^{abc}$ are the metric and structure constants in the
ordinary 3-algebra. The invariance of them demands respectively that $%
f^{abcd}=f_{e}^{abc}h^{ed}~$is totally antisymmetric and $f_{d}^{abc}$%
~satisfies the fundamental identity. $V_{m}^{ab}$ is antisymmetric in $a,b,$
and $C_{m,n}$ is antisymmetric in $m,n,$ while $h^{ab}$ is symmetric in $a,b$%
.~From (\ref{alge03}), we see that $g_{mn,lp}$ is antisymmetric in $m,n$.

We need to check the Jacobi identities for the above assumed algebra. We
first check the identity:
\begin{equation}
\left\langle (V_{m}^{ab},T_{n}^{c})~,T_{l}^{d}\right\rangle -\left\langle
(V_{m}^{ab},T_{l}^{d})~,T_{n}^{c}\right\rangle =[V_{m}^{ab},\left\langle
T_{n}^{c}~,T_{l}^{d}\right\rangle ]
\end{equation}%
If we use the relations (\ref{alge01}),(\ref{alge02}),(\ref{alge03}), we
find that the above is equivalent to\vspace{1pt}%
\begin{eqnarray}
\lbrack V_{m}^{ab},C_{n,l}] &=&0  \label{vc01} \\
\lbrack V_{m}^{ab},V_{n+l}^{cd}]
&=&~f_{e}^{abc}V_{m+n+l}^{ed}-~f_{e}^{abd}V_{m+n+l}^{ec}+~f^{abcd}(C_{m+n,l}+C_{m+l,n})
\label{vv01}
\end{eqnarray}%
Further if we set $l=0,~$due to antisymmetry property in $m,n$, we see that
\begin{equation}
C_{p,0}=0
\end{equation}%
for any integer $p$. So (\ref{vc01}),(\ref{vv01}) become simplified to
\begin{eqnarray}
\lbrack V_{m}^{ab},C_{n,l}] &=&0  \label{vc02} \\
\lbrack V_{m}^{ab},V_{n}^{cd}]
&=&~f_{e}^{abc}V_{m+n}^{ed}-~f_{e}^{abd}V_{m+n}^{ec}+~f^{abcd}C_{m,n}
\label{vv02}
\end{eqnarray}%
In Appendix \ref{identity}, we have shown that the $fV$ terms on the right hand side of (%
\ref{vv02}) are in fact antisymmetric under the exchange of $abm$, $cdn~$%
pairs, by virtue of the fundamental identity, and by using that $%
f_{d}^{abc}~ $furnishes a faithful and matrix representation of $%
(V_{0}^{ab})_{d}^{c}$.

We then check the Jacobi identity:
\begin{equation}
\lbrack
V_{l}^{fg},[V_{m}^{ab},V_{n}^{cd}]]=[[V_{l}^{fg},V_{m}^{ab}],V_{n}^{cd}]-[[V_{l}^{fg},V_{n}^{cd}],V_{m}^{ab}]
\label{jacobi_v}
\end{equation}%
We use the relations in (\ref{vv02}) to simplify the above. The above is
equivalent to two equations. One equation with the $V~$terms are satisfied
due to the fundamental identity for the structure constant$~f_{d}^{abc}.~$

The other equation imposes restrictions on $C_{m,n}.$
\begin{eqnarray}
&&(f_{e}^{abc}f^{fged}-f_{e}^{abd}f^{fgec})C_{l,m+n}  \notag \\
&=&(f_{e}^{fgb}f^{cdea}-f_{e}^{fga}f^{cdeb})C_{n,l+m}+(f_{e}^{fgc}f^{abed}-f_{e}^{fgd}f^{abec})C_{m,l+n}
\end{eqnarray}

The coefficients for $C_{l,m+n}$ and $C_{m,l+n}$ are negative with respect
to each other. This is because of the identities
\begin{eqnarray}
f_{e}^{abd}f^{fgec} &=&-f_{e}^{abd}f_{e^{\prime }}^{fgc}h^{ee^{\prime }}
\label{ffh01} \\
f_{e}^{fgc}f^{abed} &=&-f_{e^{\prime }}^{abd}f_{e}^{fgc}h^{ee^{\prime }}
\label{ffh02}
\end{eqnarray}%
\vspace{1pt}Thereby%
\begin{equation}
f_{e}^{abd}f^{fgec}=f_{e}^{fgc}f^{abed}  \label{ffh03}
\end{equation}%
since the metric $h^{ee^{\prime }}$ is symmetric, and similarly for another
term. The derivation in (\ref{ffh01}),(\ref{ffh02}),(\ref{ffh03}) only
assumes that the metric $h^{ab}$ is symmetric and is independent of the
signature of the metric. So this holds for any signature, including
Euclidean, Minkowski signatures and the case when there are zero eigenvalues
in the metric.

The coefficients for $C_{l,m+n}$ and $C_{n,l+m}~$are also negative with
respect to each other, and this is because
\begin{eqnarray}
&&f_{e}^{abc}f^{fged}-f^{fgec}f_{e}^{abd}-f_{e}^{fga}f^{cdeb}+f_{e}^{fgb}f^{cdea}
\notag \\
&=&(f_{e}^{abc}f_{d^{\prime }}^{fge}-f_{e}^{fgc}f_{d^{\prime
}}^{abe}-f_{e}^{fga}f_{d^{\prime }}^{bce}-f_{e}^{fgb}f_{d^{\prime
}}^{cae})h^{dd^{\prime }}=0
\end{eqnarray}%
This is zero since it is the fundamental identity in the bracket contracted
with the metric. In the above derivation we also only used that the metric
is symmetric and the derivation is independent of the signature of the
metric.

Thereby we have
\begin{equation}
C_{l,m+n}+C_{n,l+m}+C_{m,n+l}=0
\end{equation}%
This relation implies the recursion relation%
\begin{equation}
C_{m,k-m}=C_{m-1,k-m+1}+C_{1,k-1}  \label{recursion01}
\end{equation}%
Using this we get $C_{2,k-2}=2C_{1,k-1},~$and using this recursion relation (%
\ref{recursion01}) $m-1$ times we get
\begin{equation}
C_{m,k-m}=mC_{1,k-1}
\end{equation}%
Thereby
\begin{equation}
kC_{1,k-1}=0
\end{equation}%
So we have $C_{m,k-m}=m\delta _{k,0}C_{1,k-1},~$or equivalently
\begin{equation}
C_{m,n}=m\delta _{m,-n}C_{1,-1}
\end{equation}%
We see that $C_{1,-1}$ is the only independent non-zero central charge, and
we may define
\begin{equation}
C\equiv C_{1,-1}  \label{central_01}
\end{equation}

Then (\ref{vv02}) is simplified to (see also \cite{Gustavsson:2008dy}
without the $C$ term)
\begin{equation}
\lbrack
V_{m}^{ab},V_{n}^{cd}]=f_{e}^{abc}V_{m+n}^{ed}-f_{e}^{abd}V_{m+n}^{ec}+f^{abcd}m\delta _{m,-n}C
\label{vv02.5}
\end{equation}%
By using the analysis in Appendix \ref{identity}, we can rewrite
the above in a way that is manifestly antisymmetric under the
exchange of $abm,cdn$ pairs:
\begin{eqnarray}
\lbrack V_{m}^{ab},V_{n}^{cd}] &=&\frac{1}{2}%
(f_{e}^{abc}V_{m+n}^{ed}-f_{e}^{abd}V_{m+n}^{ec}+f_{e}^{cdb}V_{m+n}^{ea}-f_{e}^{cda}V_{m+n}^{eb})+f^{abcd}m\delta _{m,-n}C
\notag  \label{vv02.6} \\
&&
\end{eqnarray}%
(\ref{vv02.5}) and (\ref{vv02.6}) are equivalent modulo the
fundamental identity, see Appendix \ref{identity}. We then need to
check (\ref{jacobi_v}) for the new expression (\ref{vv02.6}), and
we find that the equation with the $C$ terms yields the same
equation, and the equation with the $V$ terms again satisfies, by
using the fundamental identity multiple times.

Now we see that $g_{mn,lp}~$and~(\ref{alge03}) are simplified to
\begin{eqnarray}
g_{mn,lp} &=&m\delta _{m,-n}~g_{l,p} \\
(C,T_{l}^{c}) &=&g_{l,p}T_{p}^{c}
\end{eqnarray}%
where $g_{l,p}$ is a function of $l$ and $p$. We have not assumed any
symmetry property for $g_{l,p}$.

We next look at the Jacobi identity:
\begin{equation}
\left\langle (C,T_{n}^{c})~,T_{l}^{d}\right\rangle -\left\langle
(C,T_{l}^{d})~,T_{n}^{c}\right\rangle =[C,\left\langle
T_{n}^{c}~,T_{l}^{d}\right\rangle ]
\end{equation}
This identity is equivalent to two equations, one for the $V$ terms, and
another for the $C$ terms:
\begin{eqnarray}
g_{n,p}V_{p+l}^{cd}+g_{l,q}V_{q+n}^{cd} &=&0 \\
g_{n,p}\delta _{p,-l}p-g_{l,q}\delta _{q,-n}q &=&0
\end{eqnarray}%
where $p~$or $q$ is summed over. By just looking at the case $l=n~$for the
first equation, we infer%
\begin{equation}
g_{l,p}=0
\end{equation}

So far, the rest of the Jacobi identities involve two elements in $\mathcal{%
B,~}$and one elements in $\mathcal{A,~}$and is equivalent to an identity of
five elements in $\mathcal{A.~}$This equation, in the present case, is the
fundamental identity for the 3-bracket algebra (\ref{alge04}), and since $%
g_{l,p}=0,$ or $g_{mn,lp}=0$, this is the same as the fundamental identity
for the structure constant $f_{d}^{abc}.~$

To summarize, the extension with mode numbers, under various consistency
conditions\footnote{%
More analysis on the manifest antisymmetry under the exchange of $abm$, $cdn$
pairs in (\ref{vv03}) is in Appendix \ref{identity}. We used expression (\ref%
{vv02.6}) instead of (\ref{vv02.5}) in (\ref{vv03}).} and assuming the
ansatz (\ref{alge01})-(\ref{alge04}), is
\begin{eqnarray}
\left\langle T_{m}^{a},T_{n}^{b}\right\rangle &=&V_{m+n}^{ab}+h^{ab}m\delta
_{m,-n}C  \label{infinite_alg_01} \\
(V_{m}^{ab},T_{n}^{c})~ &=&~f_{d}^{abc}T_{m+n}^{d} \\
(C,T_{l}^{c}) &=&0 \\
\lbrack V_{m}^{ab},V_{n}^{cd}] &=&\frac{1}{2}%
(f_{e}^{abc}V_{m+n}^{ed}-f_{e}^{abd}V_{m+n}^{ec}+f_{e}^{cdb}V_{m+n}^{ea}-f_{e}^{cda}V_{m+n}^{eb})
\notag \\
&&+f^{abcd}m\delta _{m,-n}C  \label{vv03} \\
\lbrack V_{m}^{ab},C] &=&0 \\
\lbrack T_{m}^{a},T_{n}^{b},T_{l}^{c}] &=&~f_{d}^{abc}T_{m+n+l}^{d}
\label{infinite_alg_06}
\end{eqnarray}%
%
%
%
%
%
%
%
%
%
%
%
%
%
%

This algebra has various subalgebras. If we look at the generators with zero
modes, i.e. if we truncate the algebra keeping only the modes $m,n,l=0,~$we
get the ordinary 3-algebra. This extended algebra of course includes the
infinite dimensional extension of the $so(4)~$3-algebra and the direct sum
of the $so(4)~$3-algebras, by adding mode numbers to each generators. The
central charge $C$ appears on the right hand sides of (\ref{vv03}) and (\ref%
{infinite_alg_01}), and may introduce normal ordering issues in the products
of operators.

If we start from (\ref{vv03}), we can look at the subalgebra by fixing $%
a=c=\ast ,$ where $\ast $ is a specified gauge index, we get (see also \cite%
{Gustavsson:2008dy})
\begin{equation}
\lbrack V_{m}^{\ast b},V_{n}^{\ast d}]=~f_{e}^{\ast bd}V_{m+n}^{\ast e}
\label{lie_subalg_01}
\end{equation}%
which is a Lie algebra, and the Jacobi identity for the 3-index structure
constant $f_{e}^{\ast bd},$ that is$~f_{e}^{\ast \lbrack bd}f_{h}^{g]e\ast
}=0$,$~$is a component equation of the fundamental identity for the 4-index
structure constant $f_{e}^{abd},$ and is satisfied as long as the
fundamental identity is satisfied.

Under the truncation (\ref{lie_subalg_01}), the central charge $C~$%
disappears on the right hand side, due to the total antisymmetry of $%
f^{abcd}~$in the last term of (\ref{vv03}), thereby this extension is not
equivalent to the usual infinite dimensional extension of Lie algebras with
central charges, and is intrinsically 3-algebraic. This also means that the
effects of $C$ may not be seen after taking the limit to a D2-brane gauge
theory. We also mention if we hypothetically had a term
\begin{equation}
g^{abcd}m\delta _{m,-n}C~~  \label{g_abcd_00}
\end{equation}%
where
\begin{equation}
g^{abcd}=\ h^{bc}h^{ad}-h^{ac}h^{bd}
\end{equation}%
on the right hand side of (\ref{vv03}), we could have kept the $C$ charge on
the right hand side of (\ref{lie_subalg_01}), but this term (\ref{g_abcd_00}%
) will not satisfy the Jacobi identities, primarily due to that $g^{abcd}$
is not totally antisymmetric, in contract with $f^{abcd}.$

\subsection{Infinite dimensional extensions with a Lorentzian metric}

\label{lorentzian extension} 

In this subsection, we discuss a different infinite dimensional extension of
the 3-algebra, that is different from the ansatz (\ref{alge01})-(\ref{alge04}%
) used in subsection \ref{infinite dimensional algebra_general metric}, and
we focus on the algebra with a metric of Minkowski or Lorentzian signature.

If we consider the 3-bracket algebra with a Lorentzian metric \cite%
{Gomis:2008uv},\cite{Benvenuti:2008bt},\cite{Ho:2008ei}, we may start from
the ansatz in \cite{Morozov:2008rc},
\begin{equation}
\lbrack
T^{a},T^{b},T^{c}]=~tr(T^{a})[T^{b},T^{c}]+tr(T^{b})[T^{c},T^{a}]+tr(T^{c})[T^{a},T^{b}]+T^{-}tr(T^{a},[T^{b},T^{c}])
\label{3-bracket_08}
\end{equation}%
where $T^{-}~$is a central element in the 3-bracket algebra. This ansatz
will be equivalent to \cite{Gomis:2008uv},\cite{Benvenuti:2008bt},\cite%
{Ho:2008ei} if we single out an identity matrix $\frac{1}{N}\mathbf{1}$ and
make other $T^{a}$s traceless.

We may directly start from a standard KM algebra for a Lie algebra,
\begin{eqnarray}
\lbrack T_{m}^{a},T_{n}^{b}] &=&\lambda _{c}^{ab}T_{m+n}^{c}+h^{ab}m\delta
_{m,-n}T^{-}  \label{km_01} \\
\lbrack T_{m}^{a},T^{-}] &=&0  \label{km_02}
\end{eqnarray}%
We can plug these 2-brackets into\footnote{%
We thank Andreas Gustavsson for making a suggestion of this different type
of extension.} the defining equation for the 3-brackets in (\ref%
{3-bracket_08}), and then we have
\begin{eqnarray}
\lbrack T_{m}^{a},T_{n}^{b},T_{l}^{c}] &=&~\lambda
_{d}^{bc}tr(T_{m}^{a})T_{n+l}^{d}+\lambda
_{d}^{ca}tr(T_{n}^{b})T_{l+m}^{d}+\lambda
_{d}^{ab}tr(T_{l}^{c})T_{m+n}^{d}+\lambda ^{abc}\delta _{m+n+l,0}T^{-}
\notag \\
&&+\{h^{bc}tr(T_{m}^{a})n\delta _{n,-l}+h^{ca}tr(T_{n}^{b})l\delta
_{l,-m}+h^{ab}tr(T_{l}^{c})m\delta _{m,-n}\}T^{-}  \label{3_bracket_km_06} \\
\lbrack T^{+},T_{m}^{a},T_{n}^{b}] &=&\lambda _{c}^{ab}tr(T^{+})T_{m+n}^{c}
\label{3_bracket_km_04} \\
\lbrack T^{-},T_{m}^{a},T_{n}^{b}] &=&0  \label{3_bracket_km_05}
\end{eqnarray}%
in which we used the metric of the 3-algebra, and $T^{+}$ is another null
generator in the Lorentzian 3-algebra. Two $h^{ab}tr(T^{+})m\delta
_{m,-n}T^{-}$ terms with opposite signs in (\ref{3_bracket_km_04}) are
cancelled. Other brackets are zero.

In this case, the fundamental identity for the 3-brackets (\ref{3-bracket_08}%
) will be satisfied, if the Jacobi identities for the 2-brackets are
satisfied \cite{Gustavsson:2008dy},\cite{Morozov:2008cb},\cite{Awata:1999dz}%
. This is indeed the case since the Jacobi identities for (\ref{km_01})-(\ref%
{km_02}) are satisfied. If we set $m=n=0$, the equation (\ref%
{3_bracket_km_04}) defines the Lie algebra where $[T^{+},\cdot ~,~\cdot ]~$%
defines the Lie algebra commutator,$~$and if we keep the general $m,n$, (\ref%
{3_bracket_km_04}) also defines an ordinary KM algebra but with the central
term disappeared, similar to the discussion in subsection \ref{infinite
dimensional algebra_general metric}.

If we make the $T_{m}^{a}$s traceless, and make a redefinition$%
~T^{+}\rightarrow kT^{+},~$where $tr(T^{+})=k\neq 0,~$then the algebra
becomes simplified:
\begin{eqnarray}
\lbrack T_{m}^{a},T_{n}^{b},T_{l}^{c}] &=&\lambda _{c}^{ab}\delta
_{m+n+l,0}T^{-}  \label{lorentzian_km_01} \\
\lbrack T^{+},T_{m}^{a},T_{n}^{b}] &=&\lambda _{c}^{ab}T_{m+n}^{c}
\label{lorentzian_km_02} \\
\lbrack T^{-},T_{m}^{a},T_{n}^{b}] &=&0  \label{lorentzian_km_03}
\end{eqnarray}%
All 6 types of fundamental identities are satisfied. This can be viewed as
an infinite dimensional extension of the Lorentzian algebra \cite%
{Gomis:2008uv},\cite{Benvenuti:2008bt},\cite{Ho:2008ei}, (\cite{Lin:200804},%
\cite{Morozov:2008rc},\cite{Gustavsson:2008dy}), and reduces to the latter
when keeping $m=n=l=0$. The $T^{-}$ generator in the Lorentzian algebra can
thereby have an interpretation of a central term in a underlying KM algebra (%
\ref{km_01}).

\vspace{1pt} \vspace{1pt} \vspace{5pt}

We may look for a generating function for the generators with different
modes, for example, if we look at the $T_{m}^{a}$ generators, we may derive
them from the expansion of
\begin{equation}
T^{a}(\sigma )=\frac{1}{2\pi }\sum_{m}T_{m}^{a}e^{im\sigma }
\label{generating_funtion_01}
\end{equation}%
where $\sigma $ is periodic with periodicity $2\pi ,$ and $m\in \mathbf{Z}%
\mathcal{.}$~The generating functions $T^{a}(\sigma )$ may be viewed as
valued in an internal direction $\sigma $.~This may be relevant if the world
volume fields on the multiple M2-branes carry not only gauge indices and
Lorentz indices, but also internal indices corresponding to boundary lines
of open membranes stretching between M2-branes. This is also relevant for
the explanation of the M2 to D2 reduction. In the above assumption, this
limit 
may involve integrating the $\sigma ~$circle, when one of the transverse
scalars has an abelian component which under a gauge choice is identified
with the $\sigma ~$circle, and receives a periodicity and is then integrated
out. 
The above assumption seems to be rather natural in explaining the appearance
of the periodicity.

It would be interesting to understand the relevance and the problem of the
classification of the physical unitary representations of such algebras,
especially the one for the $so(4)~$3-algebra and the direct sum of the $%
so(4)~$3-algebras, as well as the Lorentzian 3-algebra. In subsection \ref%
{lagrangian}, we also emphasize that a Lagrangian 2-tensor naturally lives
in the algebra of $\mathcal{B}$.


\section{Extensions with different signatures of the metric}

\label{algebra-metric-lagrangian}


\subsection{Algebras with different signatures}

\label{algebra_metric} 

In this section, we first study the algebra with different signatures of the
metric, with the motivation of embedding a general Lie algebra, including
the case of semisimple Lie algebras and the case of their direct sum with
abelian ones. We consider both the cases when the metric has a negative
eigenvalue and when it has a zero eigenvalue.

If we want to form a Lie subalgebra, we may pick a index $+$, similar to the
relation in (\ref{lie_subalg_01}) when we pick a index $\ast $, so that
\begin{equation}
f_{c}^{+ab}=\lambda _{c}^{ab}  \label{f+abc_01}
\end{equation}%
where $\lambda _{c}^{ab}$ is a Lie algebra structure constant.

In this case, the covariant derivative
\begin{equation}
D_{\mu }X_{a}^{I}=\partial _{\mu }X_{a}^{I}-f^{dbc}{}_{a}A_{\mu cd}X_{b}^{I}
\end{equation}
contains a piece
\begin{equation}
\partial _{\mu }X_{a}^{I}-\lambda _{a}^{bc}{}A_{\mu c}^{\prime }X_{b}^{I}
\end{equation}
where $A_{\mu c}^{\prime }=2A_{\mu c+},~$which looks the same as in a
conventional gauge theory.

However we do not want $T^{+}~$to appear also on the right hand side of the
3-brackets, since if that is the case we will have a very strong Plucker
type relation
\begin{equation}
\lambda ^{cde}\lambda _{g}^{ab}=\lambda ^{ab[c}\lambda _{g}^{de]}
\label{plucker_type_01}
\end{equation}%
from the fundamental identity, when checking $%
[T^{a},T^{b},[T^{c},T^{d},T^{e}]]$. This identity will only allow $so(3),$
direct sum of $so(3)$s, and the direct sum of them with $u(1)$s, as
solutions. We want to avoid this identity so we let
\begin{equation}
f_{+}^{abc}=0
\end{equation}

Then we need to check the total antisymmetry of $f^{+abc}:$
\begin{eqnarray}
f^{+abc} &=&f_{c}^{+ab}=\lambda _{c}^{ab} \\
f^{abc+} &=&f_{+}^{abc}h^{++}+f_{-}^{abc}h^{-+}  \notag \\
&=&f_{-}^{abc}h^{-+}=-\lambda _{c}^{ab}  \label{f_abc+_02}
\end{eqnarray}%
where we used that the metric in the Lie algebra subspace is Euclidean.

Since $f_{-}^{abc}h^{-+}$ is non-zero, we infer that we must pick another
generator $T^{-}$~which has mixing with $T^{+}$ in the metric. We want to
check the total antisymmetry of $f^{-abc}:$
\begin{eqnarray}
f^{-abc} &=&f_{c}^{-ab} \\
f^{abc-} &=&f_{-}^{abc}h^{--}+f_{+}^{abc}h^{-+}  \notag \\
&=&f_{-}^{abc}h^{--}=-f_{c}^{-ab}  \label{f_abc-_01}
\end{eqnarray}%
We can rotate the subspace of $T^{-}$~and $T^{+},$ so there is no need to
put $f_{c}^{-ab}$ as another copy of the Lie algebra structure constant,
since we can redefine $T^{-}$~and $T^{+}$ by $T^{-}-T^{+}~$and $T^{-}+T^{+}$%
. Because of this symmetry, we can choose that $f_{c}^{+ab}$ gives the Lie
algebra structure constant, while making
\begin{equation}
f_{c}^{-ab}=0
\end{equation}

From the first derivation in (\ref{f_abc+_02}) we know $f_{-}^{abc}\neq 0,$%
~thereby from (\ref{f_abc-_01}) we see
\begin{equation}
h^{--}=0
\end{equation}

Without loss of generality we can choose
\begin{eqnarray}
f_{-}^{abc} &=&\lambda _{c}^{ab} \\
h^{-+} &=&-1
\end{eqnarray}%
from (\ref{f_abc+_02}). If we choose opposite signs for $f_{-}^{abc}$ and$%
~h^{-+},$ this would be equivalent to redefining $T^{-}$ as $-T^{-},$ so
this sign option is not necessary.

The total antisymmetry of the$~f^{ab+-}=0~$is trivially satisfied in this
algebra, and we have assumed that there is no mixing of metric between the $%
+,-$ subspace and the $a,b$ subspace.

We look at the determinant of the metric
\begin{equation}
\mathrm{det}~h=h^{--}h^{++}-(h^{-+})^{2}=-1
\end{equation}

Now we look at the value of $h^{++}.$ The value of it will not change the $%
\mathrm{det}~h=-1$. Thereby there is still a symmetry.\ This value can be
shifted away by redefining $T^{+}~$as
\begin{equation}
T^{+}+\frac{1}{2}h^{++}T^{-}
\end{equation}%
which completely fixed that symmetry. Now the new $T^{+}$ has metric%
\begin{equation}
h^{++}=0  \label{h++_01}
\end{equation}%
which is a simplified choice.

Thereby for this algebra, the bracket $[T^{+},~\cdot ~,~\cdot ]~$defines the
Lie algebra commutator. The fundamental identity is satisfied due to the
Jacobi identity of the Lie algebra structure constant, which is the only
non-trivial identity for this case. This algebra has been obtained
independently by \cite{Gomis:2008uv},\cite{Benvenuti:2008bt},\cite{Ho:2008ei}
and independently by the author \cite{Lin:200804} before the appearance of
\cite{Gomis:2008uv},\cite{Benvenuti:2008bt},\cite{Ho:2008ei}. In the above,
we present a modest derivation, with the new emphasis that this embedding is
a very rare solution to the fundamental identity and does not admit obvious
alternatives. The above eq. (\ref{plucker_type_01}) would also imply that we
can add at most products of $so(3)$s or abelian ones to the Lorentzian
3-algebra. The above derivation also makes a preparation for the discussion
below in the case of a zero eigenvalue in the metric.

Now we discuss the situation when there is a zero eigenvalue in the metric,
for example if the metric has the signature $(0,+,+,...,+).$ We denote the
null generator as $T^{0}.$ So we have
\begin{equation}
h^{00}=0,~~~~\ h^{ab}=\delta ^{ab}  \label{zero_eigen_metric_01}
\end{equation}

We want to consider the value of $f_{c}^{0ab}.$ If we make this as a
structure constant of a Lie algebra, like (\ref{f+abc_01}), then in order to
avoid the strong relation in (\ref{plucker_type_01}), we need another null
generator, which has mixing with $T^{0}~$in the metric,$~$see e.g. (\ref%
{f_abc+_02})$.$~This goes back to the $\mathrm{det}~h=-1~$case in the
previous discussion. So we would try to make simply
\begin{equation}
f_{c}^{0ab}=0
\end{equation}%
However, we can still make $f_{0}^{abc}~$as a structure constant $\lambda
_{c}^{ab}~$of a Lie algebra, without violating any constraints. Thereby we
have the simple algebra
\begin{eqnarray}
\lbrack T^{a},T^{b},~T^{c}] &=&~\lambda _{c}^{ab}T^{0}
\label{zero_eigen_alg_01} \\
\lbrack T^{0},T^{a},~T^{b}] &=&~0  \label{zero_eigen_alg_02}
\end{eqnarray}

The metric invariance is satisfied since
\begin{eqnarray}
f_{0}^{abc} &=&\lambda _{c}^{ab},~~~f_{c}^{ab0}=0,~~\ f_{d}^{abc}=0 \\
f^{abc0} &=&0,~~~~f^{abcd}=0
\end{eqnarray}%
The fundamental identity is also satisfied.

The theory corresponding to this algebra (\ref{zero_eigen_alg_01}),(\ref%
{zero_eigen_alg_02}),(\ref{zero_eigen_metric_01}) will be studied in the
second part of the subsection \ref{lagrangian}. It is much less appealing
than the $\mathrm{det}~h=-1$ case, however it has an advantage that there is
no any negative components in the metric, and the resulting theory is
manifestly unitary.

\subsection{Lagrangians with different signatures}

\label{lagrangian}

The Lagrangian of the corresponding theory was derived by first obtaining
the on-shell equations of motion, after examining the closure of the
supersymmetry algebra in \cite{Bagger:2007jr},\cite{Bagger:2007vi},\cite%
{Gustavsson:2007vu}, and later contracted with a metric. We may write the
Lagrangian in the form
\begin{equation}
\mathcal{L}=\mathcal{L}_{ab}h^{ab}
\end{equation}%
$\mathcal{L}$~must be invariant under gauge transformations. In the
component form, $\mathcal{L}_{ab}$ is \
\begin{eqnarray}
\mathcal{L}_{ij} &=&-\frac{1}{2}(\partial _{\mu }X_{i}^{I}-\tilde{A}_{\mu
}{}^{b}{}_{i}X_{b}^{I})(\partial ^{\mu }X_{j}^{I}-\tilde{A}^{\mu
}{}^{b}{}_{j}X_{b}^{I})+\frac{i}{2}\bar{\Psi}_{i}\Gamma ^{\mu }(\partial
_{\mu }\Psi _{j}-\tilde{A}_{\mu }{}^{b}{}_{j}\Psi _{b})  \notag \\
&&-\frac{1}{12}%
f_{i}^{abc}f_{j}^{efg}{}X_{a}^{I}X_{b}^{J}X_{c}^{K}X_{e}^{I}X_{f}^{J}X_{g}^{K}+%
\frac{i}{4}f_{i}^{abc}\bar{\Psi}_{b}\Gamma _{IJ}X_{c}^{I}X_{j}^{J}\Psi _{a}
\label{l_ij_01} \\
&&+\frac{1}{2}\varepsilon ^{\mu \nu \lambda }(f_{i}^{abc}A_{\mu ab}\partial
_{\nu }A_{\lambda cj}+\frac{2}{3}f_{i}^{cdb}{}f_{j}^{efa}A_{\mu ab}A_{\nu
cd}A_{\lambda ef})  \notag
\end{eqnarray}%
where we used $i,j$ indices in place of $a,b$ for clarity purpose, and the
gauge connection is $(\tilde{A}_{\mu })_{i}^{a}=(A_{\mu })_{cd}f_{i}^{cda}\,$%
. In this component form, the structure constants only appear with 3 upper
indices and 1 lower indices, and the gauge indices in the fields $%
X_{a}^{I},\Psi _{a},A_{\mu ab}$ only appear as lower gauge indices, so we
have not used the metric yet and this expression is independent of the
metric choice $h^{ij}$. So far the only assumption on the metric is that it
is symmetric and gauge invariant.

The first term in the third line of (\ref{l_ij_01}) may be replaced by the
term%
\begin{equation}
+\frac{1}{2}\varepsilon ^{\mu \nu \lambda }f_{i}^{abc}A_{\mu cj}\partial
_{\nu }A_{\lambda ab}
\end{equation}%
since they differ by a total derivative term which may not be important for
the theory defined on $R^{2,1}$ with no boundaries.

We may look at a gauge invariant 2-tensor
\begin{equation}
\mathcal{L}^{ab}=h^{ab}\mathcal{L}  \label{lagrangian-2-tensor_01}
\end{equation}%
This is gauge invariant since both $h^{ab}~$and $\mathcal{L~}$are gauge
invariant, and $\mathcal{L}^{ab}$ is an element in the algebra $\mathcal{B,}$
as discussed in section \ref{infinite dimensional algebra}. In other words,
\begin{equation}
\lbrack V_{0}^{cd},~\mathcal{L}^{ab}]=0,~~~\ [C,~\mathcal{L}^{ab}]=0
\end{equation}%
where $V_{0}^{cd}~$is $V_{m=0}^{cd}$, the zero-mode generators discussed in
section \ref{infinite dimensional algebra}, and is an arbitrary gauge
transformation.

It is interesting to note that $(\tilde{F}_{\mu \nu })_{a}^{b}$ is in the
algebra $\mathcal{B}$, and the on-shell equation of motion \cite%
{Bagger:2007jr}-\cite{Gustavsson:2007vu} relates it to
\begin{equation}
(\tilde{F}_{\mu \nu })_{a}^{b}=-\epsilon _{\mu \nu \lambda
}f_{a}^{cdb}(X_{c}^{J}D^{\lambda }X_{d}^{J}+\frac{i}{2}\bar{\Psi}_{c}\Gamma
^{\lambda }\Psi _{d})  \label{current_01}
\end{equation}%
which means that $(\tilde{F}_{\mu \nu })_{a}^{b}$ contains no new
independent degrees of freedom, and is, on-shell, the Hodge dual of the
bilinear current of the scalars and spinors. This equation is intimately
related to that the self-coupling of the gauge fields is of the Chern-Simons
type \cite{Schwarz:2004yj} or the like. Both sides of (\ref{current_01}) are
the sources coupled to the gauge fields. Moreover, since the $X_{d}^{J}$ and
$\Psi _{d}$ fields live in the algebra $\mathcal{A,}$ this equation is also
very supportive of the view that elements in $\mathcal{B}$ are formed by
anti-symmetric bilinear products of the elements in $\mathcal{A.}$

In the rest of this subsection, we discuss the theory for the case when
there is a zero eigenvalue in the metric, as in the algebra (\ref%
{zero_eigen_alg_01}),(\ref{zero_eigen_alg_02}),(\ref{zero_eigen_metric_01})
in subsection \ref{algebra_metric}. \ In this case $h^{00}\mathcal{L}_{00}=0$%
, so $\mathcal{L}_{00}$ does not contribute to the Lagrangian $\mathcal{L.}$
However, $\mathcal{L}_{00}~$has its own equations of motion. Let's discuss
the equations of motion corresponding to $\mathcal{L}_{00}=\mathcal{L}%
^{\prime }.$
\begin{eqnarray}
\mathcal{L}^{\prime } &=&-\frac{1}{2}(\partial _{\mu }X_{0}^{I}-\tilde{A}%
_{\mu }{}^{b}{}_{0}X_{b}^{I})(\partial ^{\mu }X_{0}^{I}-\tilde{A}^{\mu
}{}^{b}{}_{0}X_{b}^{I})+\frac{i}{2}\bar{\Psi}_{0}\Gamma ^{\mu }(\partial
_{\mu }\Psi _{0}-\tilde{A}_{\mu }{}^{b}{}_{0}\Psi _{b})  \notag \\
&&+\frac{i}{4}f_{0}^{abc}\bar{\Psi}_{b}\Gamma _{IJ}X_{c}^{I}\Psi
_{a}X_{0}^{J}-\frac{1}{12}%
f_{0}^{abc}f_{0}^{efg}{}{}X_{a}^{I}X_{b}^{J}X_{c}^{K}X_{e}^{I}X_{f}^{J}X_{g}^{K}
\label{L_00_01} \\
&&+\frac{1}{2}\varepsilon ^{\mu \nu \lambda }(\tilde{A}_{\mu
}{}^{a}{}_{0}\partial _{\nu }A_{\lambda a0}+\frac{2}{3}A_{\mu ab}\tilde{A}%
_{\nu }{}^{b}{}_{0}\tilde{A}_{\lambda }{}^{a}{}_{0})  \notag
\end{eqnarray}%
where $(\tilde{A}_{\mu })_{0}^{a}=(A_{\mu })_{cd}f_{0}^{cda}.~$Again, the
first term in the last line of (\ref{L_00_01}) can be replaced by
\begin{equation}
+\frac{1}{2}\varepsilon ^{\mu \nu \lambda }A_{\mu a0}\partial _{\nu }\tilde{A%
}_{\lambda }{}^{a}{}_{0}
\end{equation}%
up to a total derivative term.

The equations of motion are
\begin{eqnarray}
&&D^{2}X_{0}^{I}-\frac{i}{2}\bar{\Psi}_{c}\Gamma _{J}^{I}X_{d}^{J}\Psi
_{b}f_{0}^{cdb}=0\  \\
&&\Gamma ^{\mu }D_{\mu }\Psi _{0}+\frac{1}{2}\Gamma
_{IJ}X_{c}^{I}X_{d}^{J}\Psi _{b}f_{0}^{cdb}=0\  \\
&&(\tilde{F}_{\mu \nu })_{0}^{b}=-\epsilon _{\mu \nu \lambda
}(X_{c}^{J}D^{\lambda }X_{d}^{J}+\frac{i}{2}\bar{\Psi}_{c}\Gamma ^{\lambda
}\Psi _{d})f_{0}^{cdb}
\end{eqnarray}

The susy transformations and gauge transformations are
\begin{eqnarray}
&&\delta X_{0}^{I}=i\bar{\epsilon}\Gamma ^{I}\Psi _{0}\  \\
&&\delta \Psi _{0}=(\partial _{\mu }X_{0}^{I}-\tilde{A}_{\mu
}{}^{b}{}_{0}X_{b}^{I})\Gamma ^{\mu }\Gamma _{I}\epsilon -\frac{1}{6}%
X_{b}^{I}X_{c}^{J}X_{d}^{K}f_{0}^{bcd}\,\Gamma _{IJK}\epsilon \  \\
&&\delta (\tilde{A}_{\mu })_{0}^{a}=i\bar{\epsilon}\Gamma _{\mu }\Gamma
_{I}X_{c}^{I}\Psi _{d}f_{0}^{cda}\,
\end{eqnarray}%
and
\begin{eqnarray}
\delta X_{0}^{I} &=&\tilde{\Lambda}_{0}^{b}X_{b}^{I}\qquad \\
\delta \Psi _{0} &=&\tilde{\Lambda}_{0}^{b}\Psi _{b}\qquad \\
\delta (\tilde{A}_{\mu })_{0}^{b} &=&D_{\mu }\tilde{\Lambda}_{0}^{b}\
\end{eqnarray}%
where $\tilde{\Lambda}_{0}^{a}=\Lambda _{cd}f_{0}^{cda}.$

We may view that $\mathcal{L}^{\prime }$ gives a certain theory by itself,
which is a gauge theory with the gauge connection $(\tilde{A}_{\mu
})_{0}^{a}.$ $\mathcal{L}^{\prime }$ is gauge invariant under the gauge
transformation corresponding to this connection. This theory is decoupled
with the theory given by $\mathcal{L},$ which is
\begin{equation}
\mathcal{L}=-\frac{1}{2}\partial _{\mu }X^{aI}\partial ^{\mu }X_{a}^{I}+%
\frac{i}{2}\bar{\Psi}^{a}\Gamma ^{\mu }\partial _{\mu }\Psi _{a}
\label{L_global}
\end{equation}%
These are free theories, with global symmetry given by the Lie algebra
associated with $f_{0}^{abc}.$ The susy and local gauge transformations are
respectively\qquad $\ $
\begin{equation}
\delta X_{a}^{I}=i\bar{\epsilon}\Gamma ^{I}\Psi _{a}\ ,~~\delta \Psi
_{a}=\partial _{\mu }X_{a}^{I}\Gamma ^{\mu }\Gamma _{I}\epsilon \qquad
\end{equation}%
\begin{equation}
\delta X_{a}^{I}=0,~~\delta \Psi _{a}=0\qquad
\end{equation}%
since $\tilde{\Lambda}_{b}^{a}=0.$

It has global symmetry transformations associated with the Lie algebra,
\begin{eqnarray}
\delta X_{a}^{I} &=&\overline{\Lambda }_{a}^{b}X^{I}\qquad \\
\delta \Psi _{a} &=&\overline{\Lambda }_{a}^{b}\Psi _{b}\
\end{eqnarray}%
where $\overline{\Lambda }_{a}^{b}$ is a global gauge transformation
parameter.

This theory is not very appealing since it is a free theory with a
Lagrangian $\mathcal{L}$ and a global symmetry, decoupled from another
theory with a Lagrangian $\mathcal{L}^{\prime }~$and a local gauge symmetry,
albeit an abelian one. However the advantage is that there is no any
negative metric component in the algebra and the theory is straightforwardly
unitary.

We also remark that if $X_{a}^{I}$ receives a vev, then $\tilde{A}_{\mu
}{}^{a}{}_{0}$ gets a mass, and after integrating out this massive gauge
field (similar to \cite{Mukhi:2008ux} or \cite{Distler:2008mk},\cite%
{Lambert:2008et},\cite{VanRaamsdonk:2008ft}), one obtains a dynamical
Yang-Mills type term of the form, from (\ref{L_00_01})
\begin{equation}
-\frac{1}{4}F_{\nu \lambda 0}^{a}F_{a0}^{\nu \lambda }
\end{equation}%
which is however abelian.

\vspace{1pt}\vspace{1pt}

\section{Conclusions and discussion}

\label{discussion}

We constructed infinite dimensional 3-algebras (\ref{infinite_alg_01})-(\ref%
{infinite_alg_06}) corresponding to extending ordinary 3-algebras by adding
mode numbers. The consistency conditions and Jacobi identities single out a
unique central charge (\ref{central_01}) that appears on the right hand side
of the algebraic relations (\ref{vv03}),(\ref{infinite_alg_01}). This may
introduce new normal ordering issues in operator products. This effect may
not be seen after the limit when the theory goes to D2-brane gauge theory,
since this centrally-extended algebra is intrinsically 3-algebraic. We also
present a different infinite dimensional extension (\ref{lorentzian_km_01})-(%
\ref{lorentzian_km_03}) for the Lorentzian 3-algebras, and interpret one of
the null generators as a central term in a underlying KM algebra (\ref{km_01}%
)-(\ref{km_02}). These extended generators may be expanded by the generating
functions like (\ref{generating_funtion_01}). It would be nice to understand
the relevance and the relation of the extended algebras with M2-branes,
especially the open membranes or wrapped membranes, and the problem of the
classification of unitary representations of these algebras, especially the
ones with $so(4)~$3-algebra and their direct sums, as well as the Lorentzian
3-algebras.

We also explored ordinary 3-algebras with different signatures of the
metric, that is consistent with metric invariance and the fundamental
identity. We revisited the algebras with a negative eigenvalue in the
metric, (\ref{f+abc_01})-(\ref{h++_01}), which were obtained by the authors
of \cite{Gomis:2008uv},\cite{Benvenuti:2008bt},\cite{Ho:2008ei},\cite%
{Lin:200804}. To avoid the problem of negative kinetic terms, we explored
the algebras with a zero eigenvalue in the metric, and present the simple
algebra in (\ref{zero_eigen_alg_01}),(\ref{zero_eigen_alg_02}),(\ref%
{zero_eigen_metric_01}). This theory is manifestly unitary, and is a local
abelian gauge theory with Lagrangian $\mathcal{L}^{\prime }$ (\ref{L_00_01})
decoupled with another global gauge theory with Lagrangian $\mathcal{L}~$(%
\ref{L_global}). We also emphasized that the Lagrangian 2-tensor $\mathcal{L}%
^{ab}~$(\ref{lagrangian-2-tensor_01}) lives naturally in the algebra $%
\mathcal{B}$, and is gauge invariant.

A particular interesting theory is the mass deformed M2-brane theory
preserving $so(4)\times so(4)$ R-symmetries, with degenerate vacua
corresponding to representations of $so(4)~$and new BPS states due to
non-central charges in the Poincare superalgebra \cite{Lin:2005nh}. The
Jacobi identity of supercharges are non-trivial as emphasized in \cite%
{Lin:2005nh},\cite{Hosomichi:2008qk},\cite{Honma:2008un} (see also the
wonderful discussions in \cite{Gomis:2008cv}), and should be checked
independently, even after obtaining the supercharge anticommutators. The
smooth 11 dimensional gravity duals of these multiple vacua \cite{Lin:2004nb}%
,\cite{Lin:2005nh} not only predicts that the vacua structure can be
described by fermion bands on a cylinder, but also that there is a duality
between $m$ fivebranes wrapping one $S^{3},$ each constructed by $n~$%
M2-branes, and $n$ fivebranes wrapping another dual $\widetilde{S}^{3},$
each constructed by $m~$M2-branes. These vacua could be viewed as fuzzy $%
S^{3}~$vacua, e.g. \cite{Guralnik:2000pb}-\cite{Berman:2006eu}, \cite%
{Basu:2004ed}. There are domain walls connecting between different $so(4)$
representations, e.g. \cite{Hosomichi:2008qk}. This is very similar to the
instantons connecting between different $so(3)$ representations in the
plane-wave matrix model, and in the gravity dual it was found \cite%
{Lin:2006tr} that when the $so(3)$ representations are very close to each
other, the tunneling is mediated by Euclidean brane processes, and in the
case when the $so(3)$ representations are not close to each other, it was
proposed \cite{Lin:2006tr} to be described by a non-perturbative tunneling
of a quantum mechanical eigenvalue system. There are also bounce solutions
studied recently \cite{Popov:2008wh}. It would be nice to understand the
tunneling between different $so(4)$ representations from a gauge theoretical
point of view, especially in the illuminating framework of \cite%
{Bagger:2007jr},\cite{Bagger:2007vi},\cite{Gustavsson:2007vu}.

\vspace{1pt}

\vspace{1pt}

\vspace{0.2cm}

\section{Acknowledgements}

It is a pleasure to thank R. O'Connell, I. Bah, K. Hanaki, J. T. Liu, J.
Maldacena, K. Turzynski for helpful discussions or communications. This work
is supported by U.S. Department of Energy under grant DE-FG02-95ER40899, and
Michigan Center for Theoretical Physics.

\vspace{1pt}

\vspace{0.35cm}

\appendix

\section{Proof of an identity and antisymmetry}

\label{identity}

In section \ref{infinite dimensional algebra_general metric} we have checked
all the Jacobi identities, the upper antisymmetry of the structure constant $%
f_{d}^{abc},~$and the symmetry of the metric $h^{ab},$~and arrived
at the general expression in
(\ref{infinite_alg_01})-(\ref{infinite_alg_06}), which is
consistent with the above mentioned three consistency conditions.
One thing remains is that (\ref{vv02.5}) is not manifestly
antisymmetric under the exchange of $abm,cdn$ pairs as in
$V_{m}^{ab},V_{n}^{cd}$, although this antisymmetry property is
obviously true for the $so(4)$ 3-algebra. In this Appendix, we
prove that this is not a problem, due to the fundamental identity.
Our analysis agrees with similar analysis and conclusion in
\cite{Gustavsson:2008dy}. We may rewrite (\ref{vv02.5}) in two
ways, which we will shown to be equivalent:
\begin{eqnarray}
\lbrack V_{m}^{ab},V_{n}^{cd}]
&=&f_{e}^{abc}V_{m+n}^{ed}-f_{e}^{abd}V_{m+n}^{ec}+f^{abcd}m\delta _{m,-n}C
\label{vvA1} \\
-[V_{n}^{cd},V_{m}^{ab}]
&=&-f_{e}^{cda}V_{m+n}^{eb}+f_{e}^{cdb}V_{m+n}^{ea}+f^{abcd}m\delta _{m,-n}C
\label{vvA2}
\end{eqnarray}%
We should understand $V_{m}^{ab}$ as operators acting on the linear
combination of the generators $T_{l}^{c}~$via the definition $%
(V_{m}^{ab},~T_{l}^{c})$ as in (\ref{alge02}). If (\ref{vvA1}),(\ref{vvA2})
are equivalent, we must have
\begin{equation}
f_{e}^{bcd}V_{m+n}^{ea}=f_{e}^{abc}V_{m+n}^{ed}+f_{e}^{acd}V_{m+n}^{eb}+f_{e}^{adb}V_{m+n}^{ec}
\end{equation}%
to be true when acting on an arbitrary linear combination of the generators,
e.g. $\alpha _{g}T_{l}^{g}.$ We then would demand
\begin{equation}
f_{e}^{bcd}(V_{m+n}^{ea},~T_{l}^{g})=f_{e}^{abc}(V_{m+n}^{ed},~T_{l}^{g})+f_{e}^{acd}(V_{m+n}^{eb},~T_{l}^{g})+f_{e}^{adb}(V_{m+n}^{ec},~T_{l}^{g})
\end{equation}%
By using (\ref{alge02}), this is simplified to%
\begin{equation}
f_{e}^{bcd}f_{h}^{eag}T_{m+n+l}^{h}=(f_{e}^{abc}f_{h}^{edg}+f_{e}^{acd}f_{h}^{ebg}+f_{e}^{adb}f_{h}^{ecg})T_{m+n+l}^{h}
\end{equation}%
This is true since the coefficients in front of $T_{m+n+l}^{h}$ form the
fundamental identity, thus this proves the equivalence of (\ref{vvA1}),(\ref%
{vvA2}) and the antisymmetry under the exchange of $abm,cdn$ pairs in $%
V_{m}^{ab},V_{n}^{cd}.~$~Q.E.D.

\end{document}